\documentclass[12pt]{article}

\begin{document}


\title{\bf
DYONS OF ONE HALF MONOPOLE CHARGE}

\author{
{\bf Rosy Teh\footnote{Permanent Address: School of Physics, Universiti Sains Malaysia, 11800 USM Penang, Malaysia}}\\
{\normalsize Institute of Particle and Nuclear Studies}\\
{\normalsize High Energy Accelerator Research Organization (KEK)}\\
\bigskip
{\normalsize Tsukuba, Ibaraki 305-0801, Japan}\\
{\bf Khai-Ming Wong}\\
{\normalsize School of Physics, Universiti Sains Malaysia}\\
{\normalsize 11800 USM Penang, Malaysia}}

\date{June 2005}
\maketitle

\begin{abstract}
We would like to present some exact SU(2) Yang-Mills-Higgs dyon solutions of one half monopole charge. These static dyon solutions satisfy the first order Bogomol'nyi equations and are characterized by a parameter, $m$. They are axially symmetric. The gauge potentials and the electromagnetic fields possess a string singularity along the negative $z$-axis and hence they possess infinite energy density along the line singularity. However the net electric charges of these dyons which varies with the parameter $m$ are finite.  
\end{abstract}


\section{Introduction}
The SU(2) Yang-Mills-Higgs (YMH) field theory in $3+1$ dimensions, with the Higgs field in the adjoint representation possess both the magnetic monopole and multimonopole solutions \cite{kn:1}-\cite{kn:4}. The 't Hooft-Polyakov monopole solution with non zero Higgs mass and self-interaction is the first monopole solution that possess finite energy. This numerical, spherically symmetric monopole solution of unit magnetic charge is invariant under a U(1) subgroup of the local SU(2) gauge group  \cite{kn:1}. 

Configurations of the YMH field theory with a unit magnetic charge are in general spherically symmetric \cite{kn:1}-\cite{kn:3} although exceptions do exit \cite{kn:5}. All other monopole configurations with magnetic charges greater than unity possess at most axial symmetry \cite{kn:4} and it has been shown that these solutions cannot possess spherical symmetry \cite{kn:6}. 

Until now, exact monopole and multimonopoles solutions exist only in the Bogomol'nyi-Prasad-Sommerfield (BPS) limit \cite{kn:3}-\cite{kn:4}. Outside of this limit, where the Higgs field potential is non-vanishing only numerical solutions exist. Asymmetric multimonopole solutions with no rotational symmetry are all numerical solutions even in the BPS limit \cite{kn:7}.

Recently, we have also shown that the extended ansatz of Ref.\cite{kn:8} possesses more exact multimonopole-antimonopole configurations. The anti-configurations of all these multimonopole-antimonopole solutions with all the magnetic monopole charges reversing their sign were also constructed \cite{kn:9}. Hence monopoles becomes antimonopoles and vice versa. Solutions with vortex rings in the presence of an antimonopole-monopole-antimonopole (A-M-A) chain were also discussed \cite{kn:10}.
We have also shown the existence of half-integer topological monopole charge solutions. This configuration possesses axial symmetry and a Dirac-like string singularity along the negative z-axis \cite{kn:5}. 
The existence of smooth Yang-Mills potentials which correspond to monopoles and vortices of one-half winding number has been demonstrated in Ref.\cite{kn:11} but no exact or numerical solutions of the YMH equations have been given.

In this paper, we would like to reexamine the axially symmetric one half monopole solutions of the Ref.\cite{kn:5} once again in more detail by introducing electric charge to the system, hence creating what is called a dyon solution with a one half monopole charge. The procedure of introducing an electric charge to a monopole solution is standard and was first shown by Julia and Zee in 1975 \cite{kn:12}.

The one half monopole that was presented in Ref.\cite{kn:5} is a particular case when the solution parameter $m=-1/2$. In this paper, we would like to show that the one half monopole solution actually possesses a parameter $m$, where $-1/2 \leq m < 0$. The Higgs field of this one half monopole solution does not possess any point zeros and hence there are no other monopoles in the configuration except the one half monopole which is located at $r=0$, where the Higgs unit vector, $\hat{\Phi}^a$, becomes indeterminate and the Higgs field is singular.

By applying Gauss' Law to the electric field of this dyon solution at large $r$, we find that the electric flux seems to come from a net electric charge of decreasing magnitude as $m$ approaches zero. When $m=0$, the dyon solution of one half monopole charge becomes the dyon of the Wu-Yang type monopole \cite{kn:8} with unit magnetic charge and zero net electric charge. 

We briefly review the SU(2) Yang-Mills-Higgs field theory in the next section and introduce the modified axially symmetric magnetic ansatz \cite{kn:8} in section 3. We also discuss the dyon solutions of one half monopole charge and their electromagnetic properties in section 3 and end with some comments in section 4.

\section{The SU(2) Yang-Mills-Higgs Theory}
\label{sect:2}
The SU(2) YMH Lagrangian in 3+1 dimensions is given by
\begin{equation}
{\cal L} = -\frac{1}{4}F^a_{\mu\nu} F^{a\mu\nu} + \frac{1}{2}D^\mu \Phi^a D_\mu \Phi^a - \frac{1}{4}\lambda(\Phi^a\Phi^a - \frac{\mu^2}{\lambda})^2, 
\label{eq.1}
\end{equation}

\noindent where $\mu$ is the Higgs field mass, $\lambda$ is the strength of the Higgs potential and $\mu/\sqrt{\lambda}$ is the vacuum expectation value of the Higgs field. The Lagrangian (\ref{eq.1}) is gauge invariant under the set of independent local SU(2) transformations at each space-time point.
The covariant derivative of the Higgs field and the gauge field strength tensor are given respectively by 
\begin{eqnarray}
D_{\mu}\Phi^{a} &=& \partial_{\mu} \Phi^{a} + \epsilon^{abc} A^{b}_{\mu}\Phi^{c},\\
\label{eq.2}
F^a_{\mu\nu} &=& \partial_{\mu}A^a_\nu - \partial_{\nu}A^a_\mu + \epsilon^{abc}A^b_{\mu}A^c_\nu.
\label{eq.3}
\end{eqnarray}
Since the gauge field coupling constant, can be scaled away, we set it to one without any loss of generality. The metric used is $g_{\mu\nu} = (- + + +)$. The SU(2) internal group indices $a, b, c$ run from 1 to 3 and the spatial indices are $\mu, \nu, \alpha = 0, 1, 2$, and $3$ in Minkowski space.

The equations of motion that follow from the Lagrangian (\ref{eq.1}) are
\begin{eqnarray}
D^{\mu}F^a_{\mu\nu} &=& \partial^{\mu}F^a_{\mu\nu} + \epsilon^{abc}A^{b\mu}F^c_{\mu\nu} = \epsilon^{abc}\Phi^{b}D_{\nu}\Phi^c,\nonumber\\
D^{\mu}D_{\mu}\Phi^a &=& -\lambda\Phi^a(\Phi^{b}\Phi^{b} - \frac{\mu^2}{\lambda}).
\label{eq.4}
\end{eqnarray}
Non-BPS solutions to the YMH theory are obtained by solving the second order differential equations of motion (\ref{eq.4}), whereas BPS solutions can be more easily obtained by solving the Bogomol'nyi equations,
\begin{equation}
B^a_i \pm D_i\Phi^a = 0,
\label{eq.5}
\end{equation}
which is of first order.

The Abelian electromagnetic field tensor as proposed by 't Hooft \cite{kn:1} is
\begin{eqnarray}
F_{\mu\nu} &=& \hat{\Phi}^a F^a_{\mu\nu} - \epsilon^{abc}\hat{\Phi}^{a}D_{\mu}\hat{\Phi}^{b}D_{\nu}\hat{\Phi}^c\nonumber\\
&=& \partial_{\mu}A_\nu - \partial_{\nu}A_\mu - \epsilon^{abc}\hat{\Phi}^{a}\partial_{\mu}\hat{\Phi}^{b}\partial_{\nu}\hat{\Phi}^c,
\label{eq.6}
\end{eqnarray}
where $A_\mu = \hat{\Phi}^{a}A^a_\mu,~~\hat{\Phi}^a = \Phi^a/|\Phi|,~~|\Phi| = \sqrt{\Phi^{a}\Phi^{a}}$. Hence the 't Hooft electric field is $E_i = F_{0i}$, and the 't Hooft magnetic field is $B_i = -\frac{1}{2}\epsilon_{ijk}F_{jk}$, where the indices, $i, j, k = 1, 2, 3$. 
The topological magnetic current, which is also the topological current density of the system is \cite{kn:13} 
\begin{equation}
k_\mu = \frac{1}{8\pi}~\epsilon_{\mu\nu\rho\sigma}~\epsilon_{abc}~\partial^{\nu}\hat{\Phi}^{a}~\partial^{\rho}\hat{\Phi}^{b}~\partial^{\sigma}\hat{\Phi}^c.
\label{eq.7}
\end{equation}
Therefore the corresponding conserved topological magnetic charge is
\begin{eqnarray}
M = \int d^{3}x~k_0 
 =  \frac{1}{4\pi} \oint d^{2}\sigma_{i}~B_i. 
\label{eq.8}
\end{eqnarray}

In the BPS limit when the Higgs potential vanishes, the energy can be written in the form 
\begin{eqnarray}
E &=& \mp\int\partial_i(B^a_i\Phi^a)~d^3 x + \int\frac{1}{2}(B^a_i \pm D_i\Phi^a)^2~d^3 x\nonumber\\
&=& \mp\int\partial_i(B^a_i\Phi^a)~d^3 x = 4\pi M\frac{\mu}{\sqrt{\lambda}},
\label{eq.9}
\end{eqnarray}
where $M$ is the ``topological charge" when the vacuum expectation value of the Higgs field, $\frac{\mu}{\sqrt{\lambda}}$, is non zero coupled with some non-trivial topological structure of the fields at large $r$.

\section{The Dyons}
\label{sect:3}
\subsection{The Ansatz}
\label{sect:3.1}
The static gauge fields and Higgs field that will lead to the dyon solutions of one half monopole charge are given respectively by, \cite{kn:8}, \cite{kn:12}
\begin{eqnarray}
A_0^a &=& \sinh\gamma(\Phi_{1}~\hat{r}^a + \Phi_{2}~\hat{\theta}^a),\nonumber\\
A_{i}^a &=& \frac{1}{r}\psi(r)\left(\hat{\theta}^{a}\hat{\phi}_i - \hat{\phi}^{a}\hat{\theta}_{i}\right)  + \frac{1}{r}R(\theta)\left(\hat{\phi}^{a}\hat{r}_{i} - \hat{r}^{a}\hat{\phi}_{i}\right),\nonumber\\
\Phi^a &=& \cosh\gamma (\Phi_{1}~\hat{r}^a + \Phi_{2}~\hat{\theta}^a),
\label{eq.10}
\end{eqnarray}
\noindent where $\Phi_1 = \frac{1}{r}\psi(r), ~\Phi_2 = \frac{1}{r}R(\theta)$ and $\gamma$ is an arbitrary constant. The spherical coordinate orthonormal unit vectors, $\hat{r}^{a}$, $\hat{\theta}^{a}$, and $\hat{\phi}^{a}$ are defined by 
\begin{eqnarray}
\hat{r}^{a} &=& \sin\theta ~\cos \phi ~\delta^a_1 + \sin\theta ~\sin \phi ~\delta^a_2 + \cos\theta \delta^a_3,\nonumber\\
\hat{\theta}^{a} &=& \cos\theta ~\cos \phi ~\delta^a_1 + \cos\theta ~\sin \phi ~\delta^a_2 - \sin\theta ~\delta^a_3,\nonumber\\
\hat{\phi}^{a} &=& -\sin \phi ~\delta^a_1 + \cos \phi ~\delta^a_2,
\label{eq.11}
\end{eqnarray}
\noindent where $r=\sqrt{x^i x_i}$, ~$\theta=\cos^{-1}(x_3/r)$, and~$\phi=\tan^{-1}(x_2/x_1)$. 
Ansatz (\ref{eq.10}) satisfies the radiation or Coulomb gauge, $\partial^i A^a_i = A^a_0 = 0$. 

\subsection{The Solutions}
\label{sect:3.2}
Upon substituting ansatz (\ref{eq.10}) into the equations of motion (\ref{eq.4}), the resulting equations are simplified to two first order differential equations,
\begin{eqnarray}
r\psi^{\prime} + \psi - \psi^2 = -m(m+1),\nonumber\\
\dot{R} + R\cot\theta - R^2 = m(m+1),
\label{eq.12}
\end{eqnarray}

\noindent where $\psi^{\prime}$ means $\frac{\partial\psi}{\partial r}$ and $\dot{R}$ means $\frac{\partial R}{\partial\theta}$. 
The solutions for $\psi$ and $R$ are then given respectively by
\begin{eqnarray}
\psi(r) &=& \frac{(m+1)-m (br)^{2m+1}}{1+(br)^{2m+1}},\nonumber\\
R(\theta) &=& (m+1)\csc\theta\left\{\cos\theta - \frac{(P_{m+1}(\cos\theta) + a~Q_{m+1}(\cos\theta))}{(P_m(\cos\theta) + a~Q_m(\cos\theta))}\right\}, 
\label{eq.13}
\end{eqnarray}
where $P_m$ and $Q_m$ are the Legendre polynomials of the first and second kind of degree $m$, respectively, and $-1/2\leq m < 0$. For solutions regular along the positive $z$-axis, we required, $R(0) = 0$, and the integration constant $a$ is set to zero. The integration constant $b$ is just a scaling factor and without any loss in generality, we set $b=1$ and the boundary conditions for $\psi(r)$ are ~~$\psi(0) = m+1$ ~~and ~~$\psi(\infty) = -m$. In these dyon solutions there are no zeros of the Higgs field and the Higgs unit vector is only indeterminate at the origin where a monopole of half unit charge is located.

From the ansatz (\ref{eq.10}), the 't Hooft gauge potential becomes,
\begin{equation}
A_{\mu} = \hat{\Phi}^{a}A^{a}_{\mu} = \frac{\sinh\gamma}{r}\sqrt{\psi(r)^2+R(\theta)^2}~\delta^0_\mu.
\label{eq.14}
\end{equation}
Hence the 't Hooft electric field is non zero when $\gamma\not=0$ and the monopole becomes a dyon. 

\subsection{The 't Hooft Magnetic Field}
\label{sect:3.3}
The 't Hooft magnetic field is however independent of the gauge fields $A^a_\mu$. To calculate for the 't Hooft magnetic field $B_i$, we rewrite the Higgs field from the spherical to the Cartesian coordinate system, \cite{kn:8} 
\begin{eqnarray}
\Phi^a &=& \Phi_{1}~\hat{r}^{a} + \Phi_{2}~\hat{\theta}^{a} + \Phi_3~\hat{\phi}^{a}\nonumber\\
&=& \tilde{\Phi}_1 ~\delta^{a1} + \tilde{\Phi}_2 ~\delta^{a2} + \tilde{\Phi}_3 ~\delta^{a3}
\label{eq.15}
\end{eqnarray}
\begin{eqnarray}
\mbox{where}~~~\tilde{\Phi}_1 &=& \sin\theta \cos \phi ~\Phi_1 + \cos\theta \cos \phi ~\Phi_2 - \sin \phi ~\Phi_3
= |\Phi|\cos\alpha \sin\beta\nonumber\\
\tilde{\Phi}_2 &=& \sin\theta \sin \phi ~\Phi_1 + \cos\theta \sin \phi ~\Phi_2 + \cos \phi ~\Phi_3
= |\Phi|\cos\alpha \cos\beta\nonumber\\
\tilde{\Phi}_3 &=& \cos\theta ~\Phi_1 - \sin\theta ~\Phi_2 = |\Phi|\sin\alpha.
\label{eq.16}
\end{eqnarray}
The Higgs unit vector can be simplified to 
\begin{eqnarray}
\hat{\Phi}^a &=& \cos\alpha \sin\beta ~\delta^{a1} + \cos\alpha \cos\beta ~\delta^{a2} + \sin\alpha ~\delta^{a3},\\
\mbox{where},~~~\sin\alpha &=& \frac{\psi\cos\theta - R \sin\theta}{\sqrt{\psi^2+R^2}},~~
\beta = \frac{\pi}{2} - \phi,\nonumber
\label{eq.17}
\end{eqnarray}
and the 't Hooft magnetic field is reduce to only the $\hat{r}_i$ and $\hat{\theta}_i$ components,
\begin{eqnarray}
B_i = -\frac{1}{r^2 \sin\theta}\left\{\frac{\partial\sin\alpha}{\partial\theta}\right\}\hat{r}_i + 
\frac{1}{r\sin\theta}\left\{\frac{\partial\sin\alpha}{\partial r}\right\}\hat{\theta}_i.
\label{eq.18}
\end{eqnarray}
The function, $\sin\alpha$, is a regular function over all space and the 't Hooft magnetic field is also regular everywhere except at the location of the monopole at the origin and along the negative $z$-axis where there is a semi infinite string singularity. When $m=-1/2$ and $m=0$, $\sin\alpha$ is independent of $r$ and the magnetic field is purely radial in direction.
We also notice that, $B_i$, can be written as
\begin{eqnarray}
B_i = \epsilon_{ijk}~\partial^j(\sin\alpha)~\partial^k\beta
= \epsilon_{ijk}~\partial^j(\sin\alpha~\partial^k\beta),
\label{eq.19}
\end{eqnarray}
and that a suitable Maxwell four-vector gauge potential for this magnetic field is
\begin{eqnarray}
{\cal A}_0 = \frac{\sinh\gamma}{r}\sqrt{\psi^2+R^2},~~~
{\cal A}_i = (\sin\alpha - 1)\partial_i\beta = -\frac{(\sin\alpha - 1)}{r\sin\theta}\hat{\phi}_i.
\label{eq.20}
\end{eqnarray}
When $m=0$, the gauge potential, ${\cal A}_i$, is just the usual Dirac string potential and it is singular along the negative z-axis. However when $-1/2 \leq m<0$, the gauge potential ${\cal A}_i$ still possesses a similar string singularity along the negative $z$-axis, but it is no longer the Dirac string.

\begin{figure}[tbh]
\vspace{5.3in}
\vskip1in
\hskip0.4in\special{eps: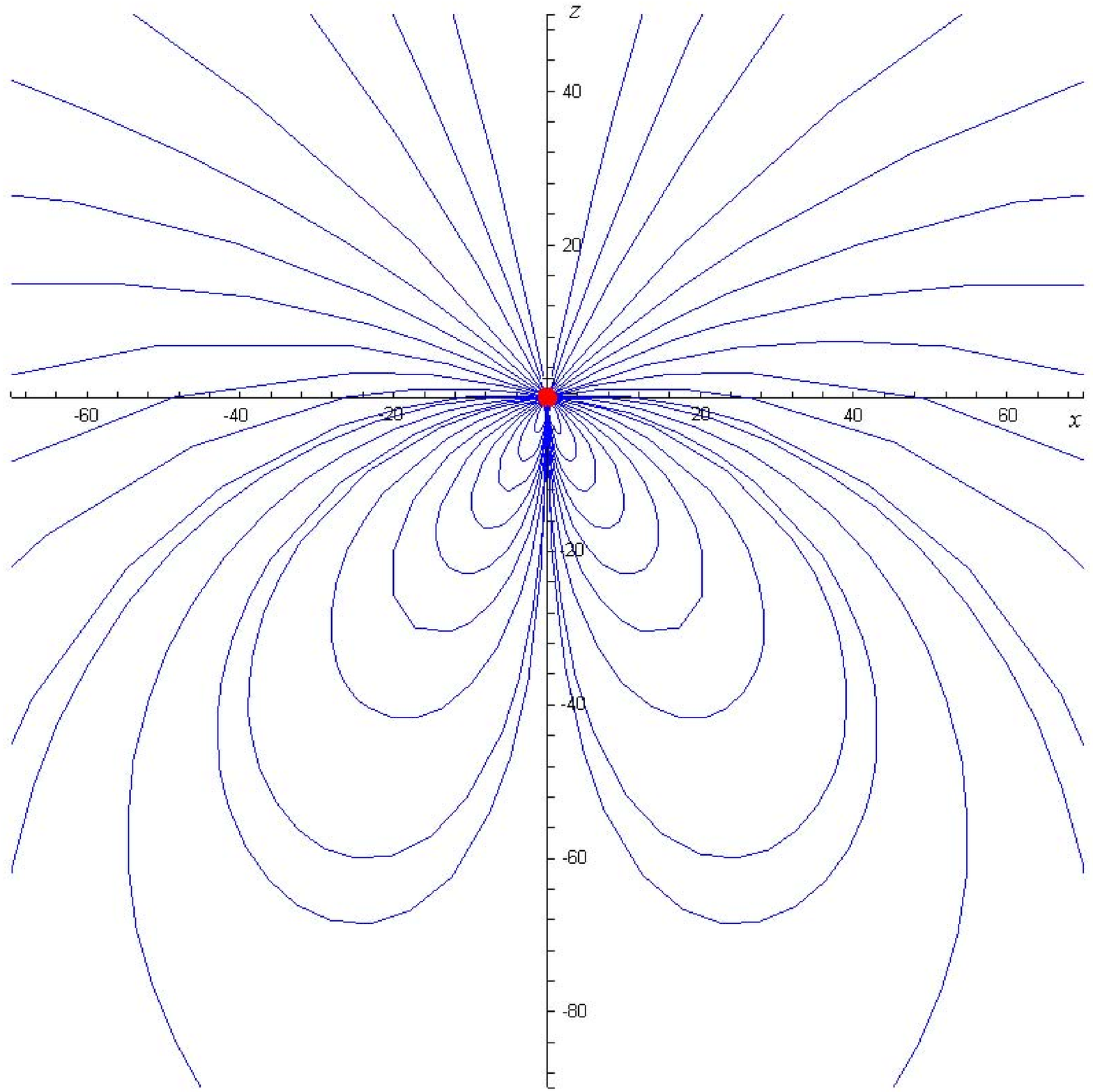 x=5.3in y=5.3in}
\caption{The magnetic field lines of the m=-0.01 dyon. The one half monopole is located at $r=0$.}
\label{fig.1}
\end{figure}

\begin{figure}[tbh]
\vspace{5.5in}
\hskip0.2in\special{eps: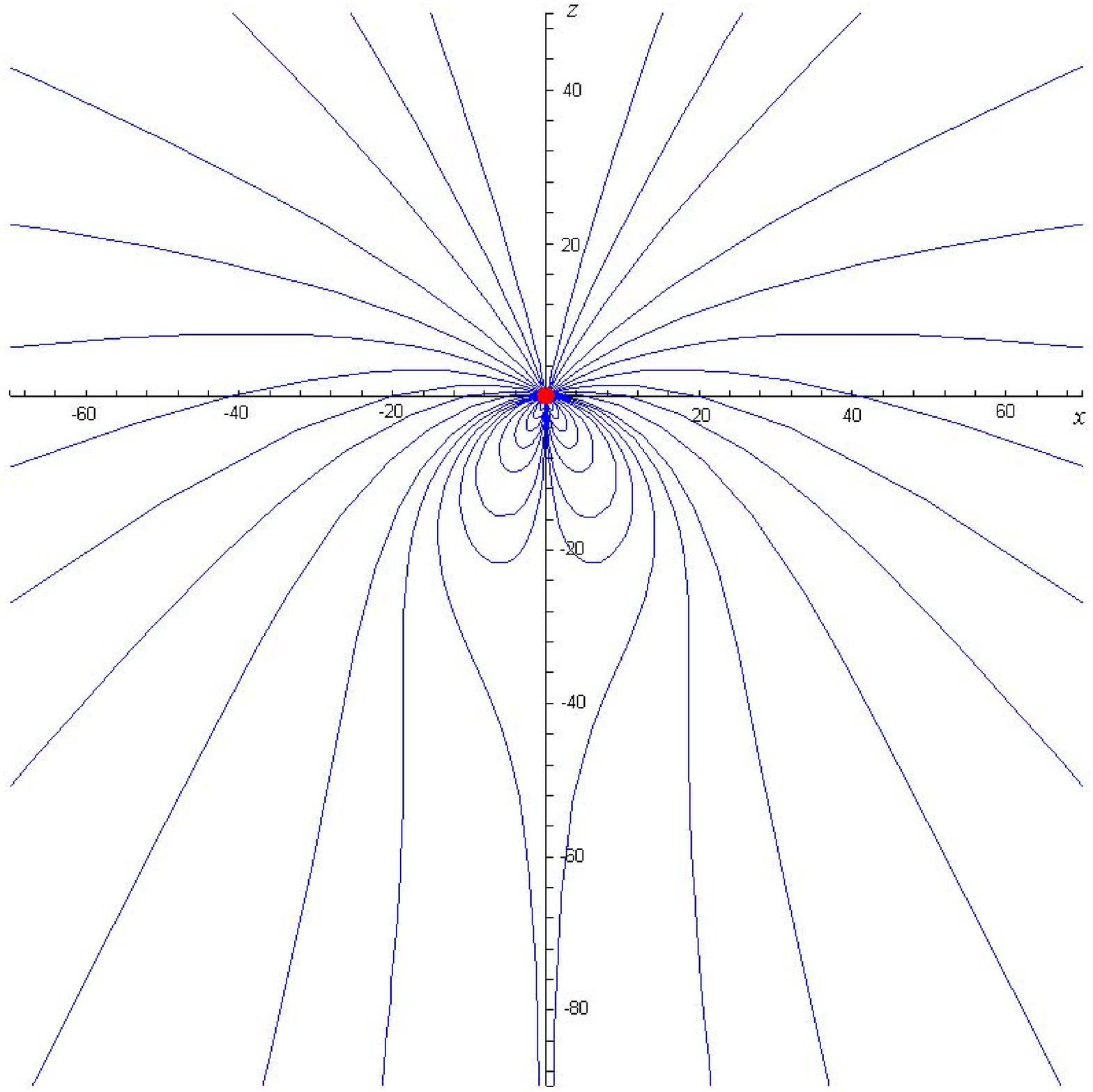 x=5.5in y=5.5in}
\caption{The magnetic field lines of the m=-0.05 dyon. The one half monopole is located at $r=0$.}
\label{fig.2}
\end{figure}

\begin{figure}[tbh]
\vspace{5.5in}
\hskip0.2in\special{eps: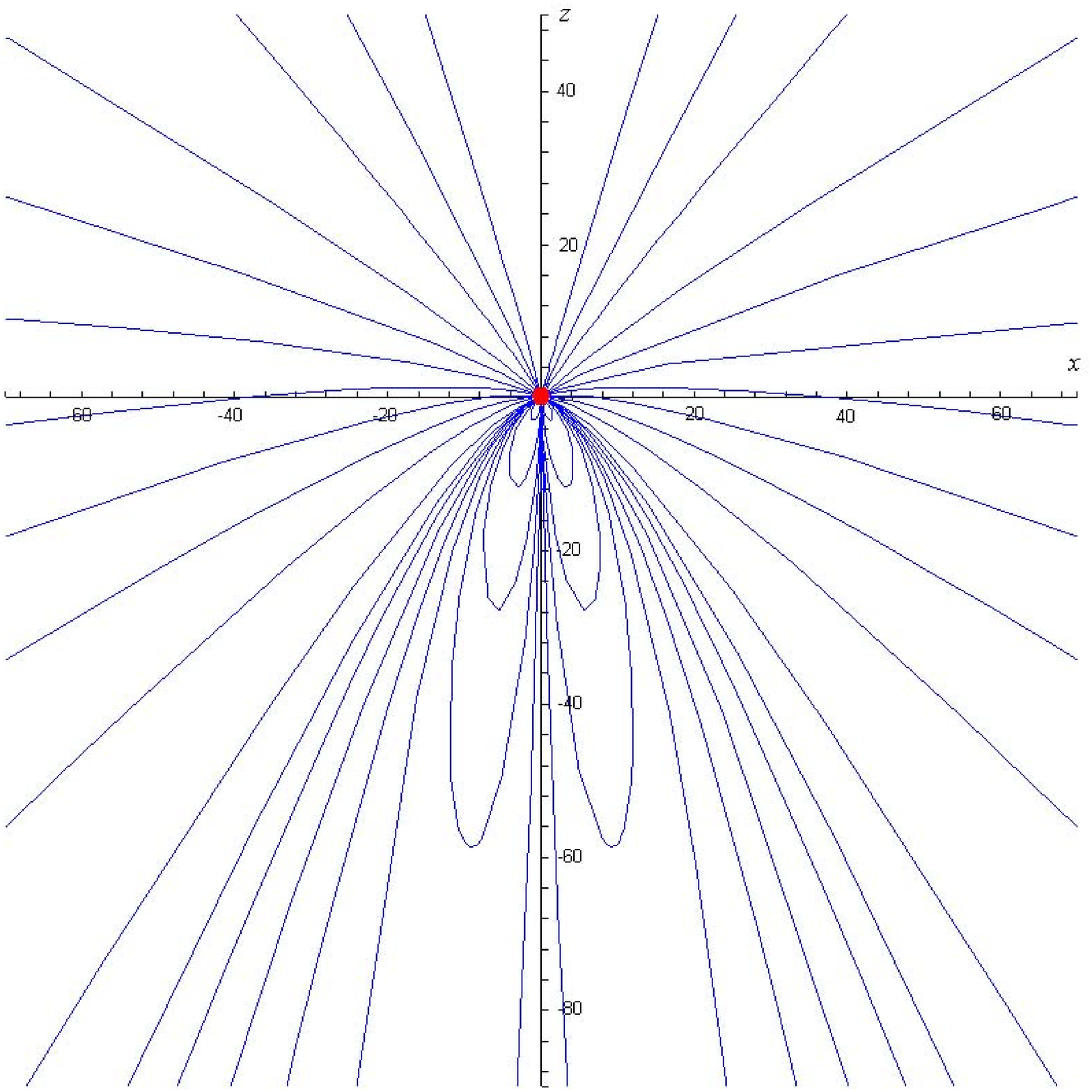 x=5.5in y=5.5in}
\caption{The magnetic field lines of the m=-0.2 dyon. The one half monopole is located at $r=0$.}
\label{fig.3}
\end{figure}

From Eq.(\ref{eq.19}), it is obvious that the magnetic field is always perpendicular to the gradients of $\sin\alpha$ and $\beta$. Hence the magnetic field lines lie on the line $\sin\alpha = k$, where $-1<k<1$, and $\phi =$ constant. By plotting $\sin\alpha = k$ on a vertical plane through the origin; we manage to draw the magnetic field lines for the configurations when $m=-0.01$, Fig.(\ref{fig.1}); $m=-0.05$, Fig.(\ref{fig.2}); and $m=-0.2$, Fig.(\ref{fig.3}). We have notice that for every field line through the origin, one end of the line goes to infinity, whereas the other end will make a loop and return back to $r=0$. Therefore the magnetic flux going to infinity is only half of that of a monopole of unit charge. 

\subsection{The Monopole Charge}
\label{sect:3.4}
From Eq.(\ref{eq.8}), the net magnetic charge enclosed by the sphere at infinity is calculated to be, 
\begin{eqnarray}
M_\infty = -\left.\frac{1}{2}\sin\alpha\right|^\pi_{0, r\rightarrow \infty} 
&=& \frac{1}{2},~~~\mbox{when}~~-\frac{1}{2}\leq m < 0,\nonumber\\
&=& 1,~~~~\mbox{when}~~m=0.
\label{eq.21}
\end{eqnarray}
Hence the magnetic charge of the system is always one half over all space when $m$ is a non integer less than zero and it is concentrated at only one point in space at $r=0$. When $m=0$, the half monopole becomes a unit monopole.
\subsection{The 't Hooft Electric Field}
\label{sect:3.5}
\begin{figure}[tbh]
\vspace{5.5in}
\hskip0.2in\special{eps: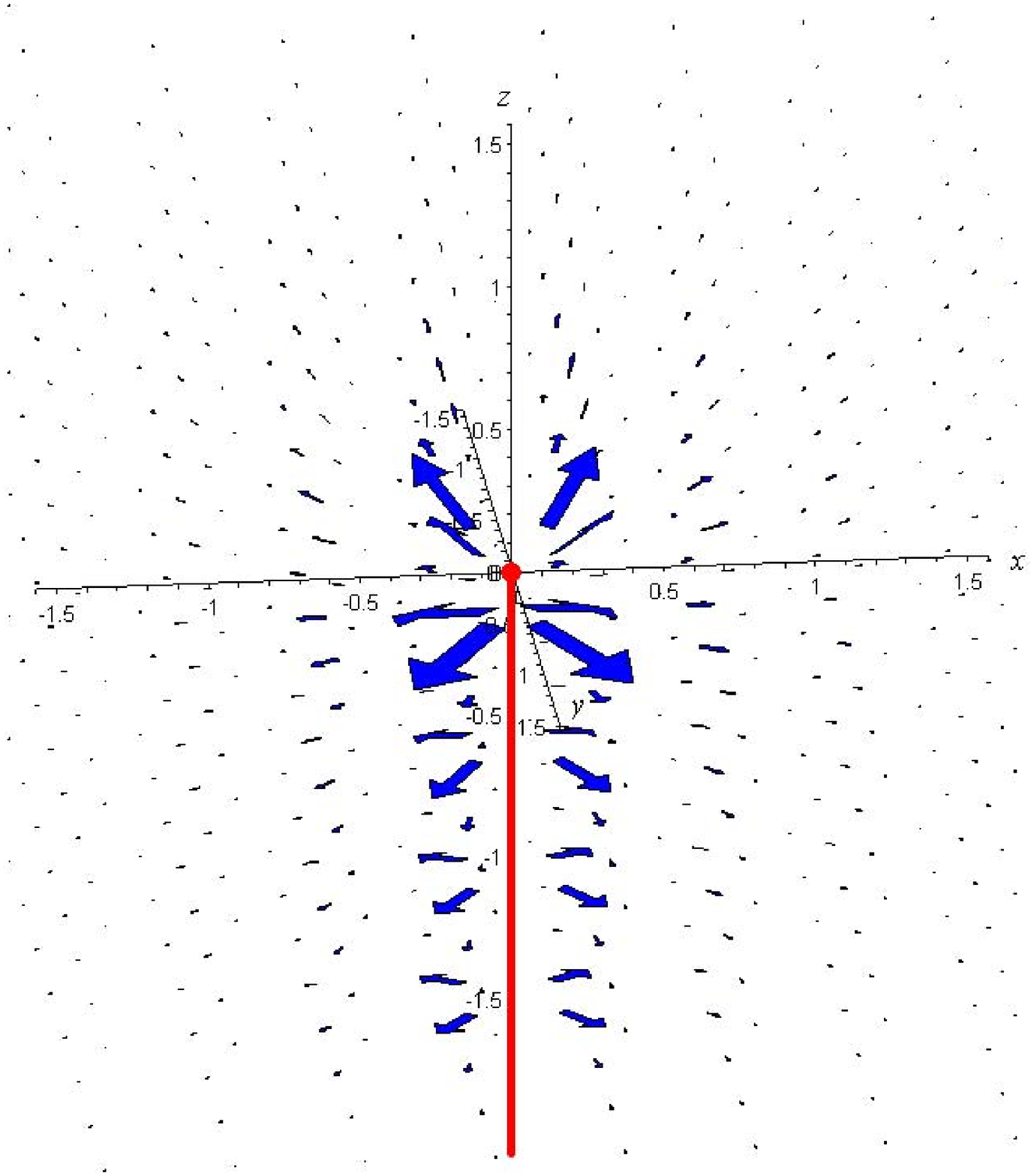 x=5.5in y=5.5in}
\caption{The electric field of the m=-0.5 dyon. The positive electric charges are concentrated at $r=0$ and along the negative $z$-axis.}
\label{fig.4}
\end{figure}

The 't Hooft electric field of the dyons is given by
\begin{eqnarray}
E_i &=& -\partial_i A_0, \nonumber\\
&=&  \sinh\gamma\left\{\sqrt{\psi^2+R^2} - \frac{\psi(\psi + m)(\psi - m-1)}{\sqrt{\psi^2+R^2}}\right\}\left(\frac{\hat{r}_i}{r^2}\right) \nonumber\\
&-& \sinh\gamma \left\{\frac{R(R^2-R\cot\theta + m(m+1))}{\sqrt{\psi^2+R^2}}\right\}\left(\frac{\hat{\theta}_i}{r^2}\right).  
\label{eq.22}
\end{eqnarray}
Unlike the magnetic field, the electric field depends on the constant $\gamma$. Hence the electric field can be switched off by setting $\gamma=0$. As the parameter $m$
decreases from $-1/2$ to zero, the $\hat{\theta}_i$ component of the electric field diminishes to zero while the radial component approaches
\begin{eqnarray}
E_i = \left\{\frac{2}{(1+r)} - \frac{1}{(1+r)^2}\right\}\frac{\hat{r_i}}{r^2},
\label{eq.23}
\end{eqnarray}
and the solution becomes spherically symmetric.
The 3D field plot of the electric field for different values of $m$ in the range $-1/2\leq m <0$, shows the presence of a positive point charge at $r=0$ and a positive line charge along the negative $z$-axis. Fig.(\ref{fig.4}) shows the 3D field plot for the case of $m=-1/2$. The electric field which is pointing radially outward diminishes as $1/r^2$ along the negative $z$-axis. 

\subsection{The Electric Charge} 
\label{sect:3.6}
\begin{figure}[tbh]
\vspace{6.7in}
\hskip0.2in\special{eps: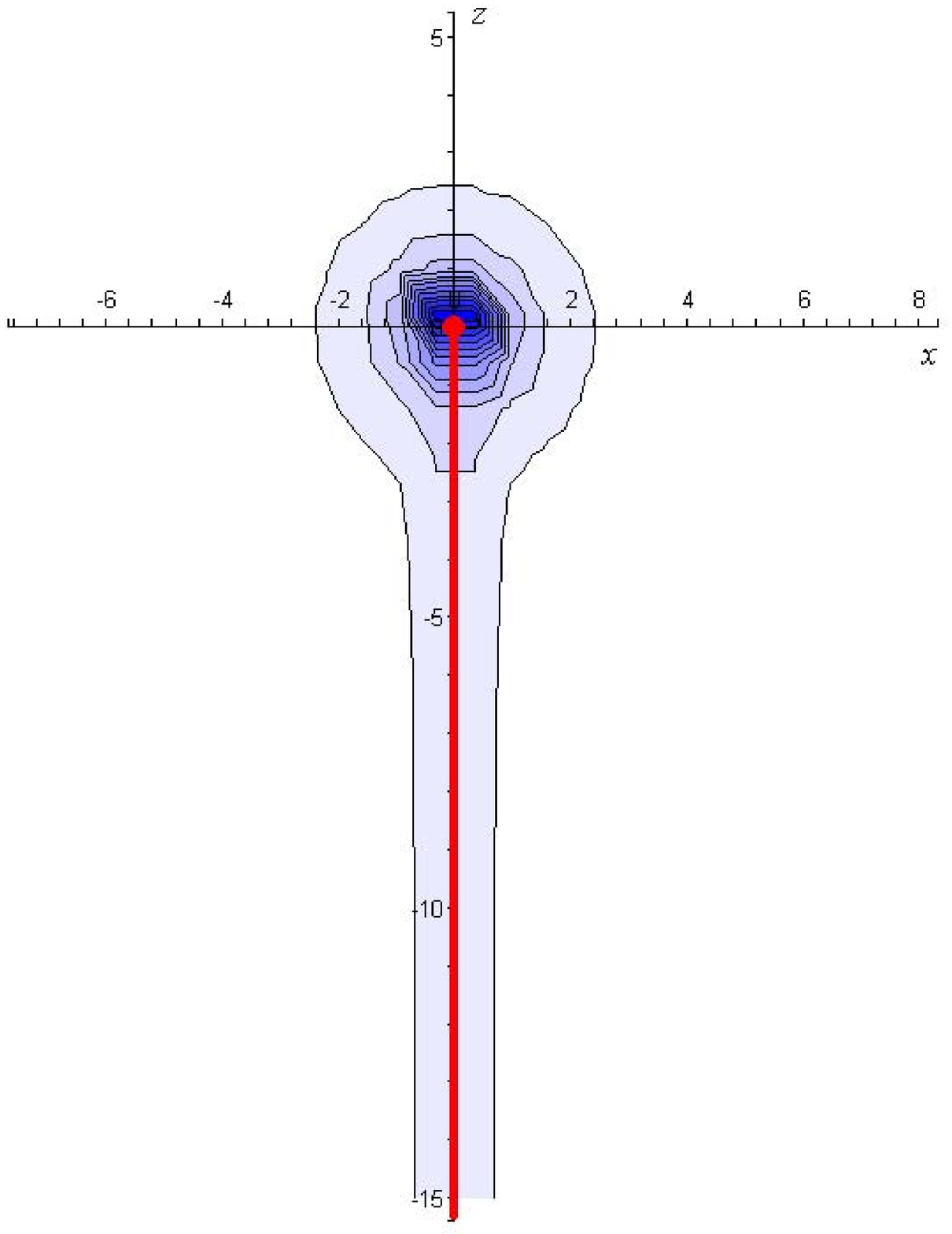 x=5in y=6.7in}
\caption{The contour plot of the electric charge density distribution of the m=-0.05 dyon. The negative electric charge density distribution is in blue and the positive charges are indicated in red.}
\label{fig.5}
\end{figure}

\begin{figure}[tbh]
\vspace{6.7in}
\hskip0.2in\special{eps: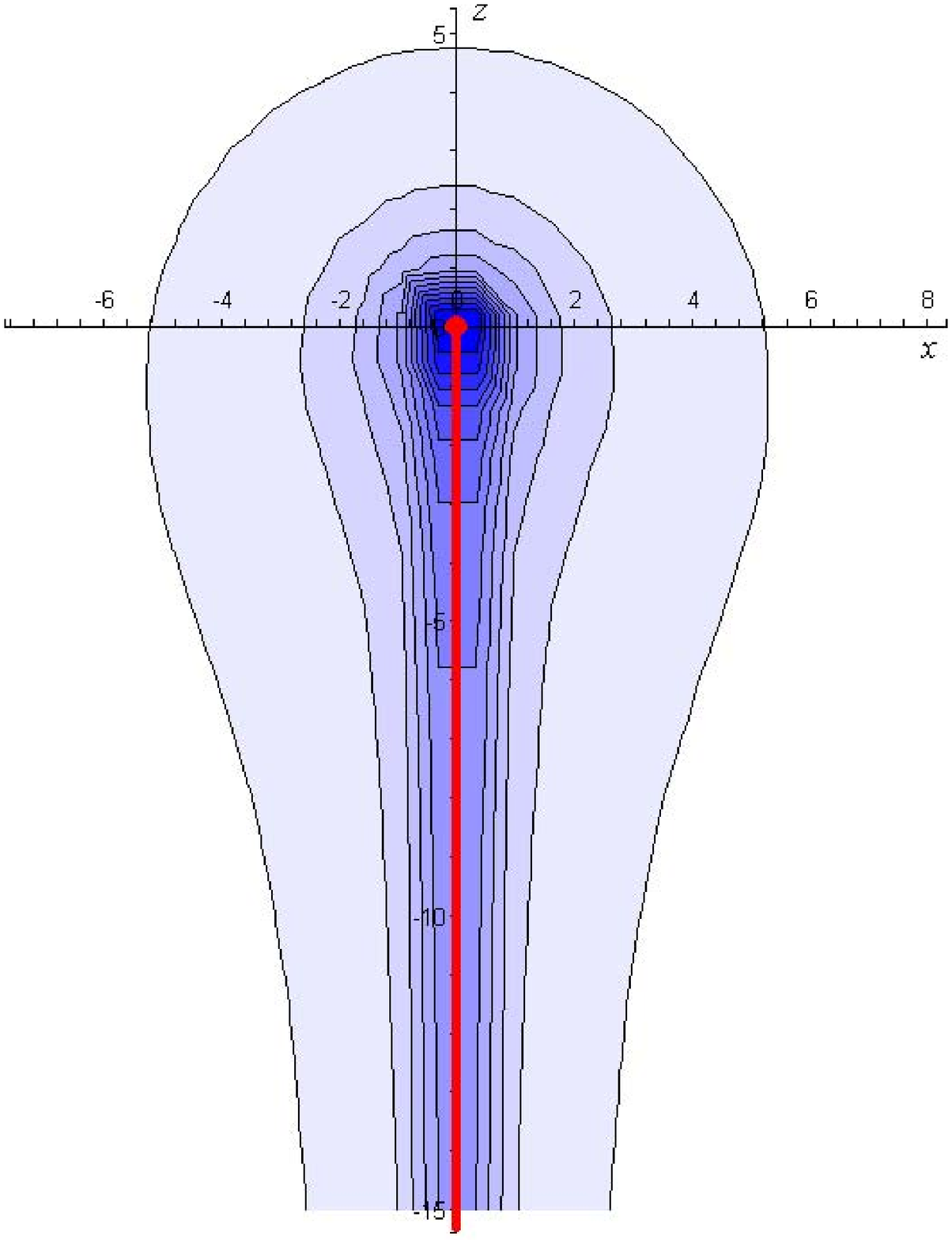 x=5in y=6.7in}
\caption{The contour plot of the electric charge density distribution of the m=-0.5 dyon. The negative electric charge density distribution is in blue and the positive charges are indicated in red.}
\label{fig.6}
\end{figure}

\begin{figure}[tbh]
\vspace{6in}
\hskip0.2in\special{eps: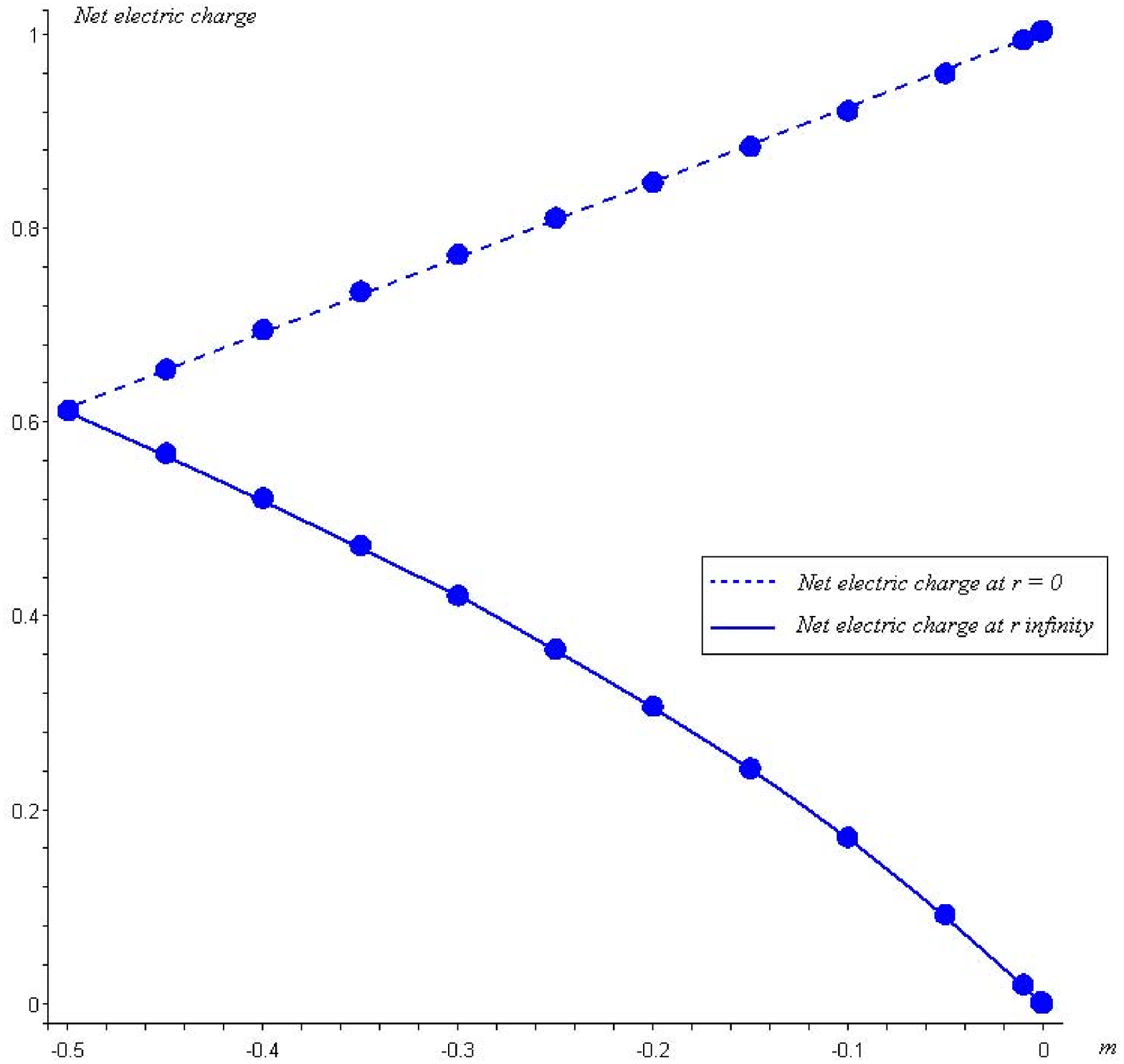 x=6in y=6in}
\caption{The point plot of the net electric charges of the dyons, $Q$ and $Q_0$ in units of $4 \pi \sinh \gamma$ versus $m$. The upper curve is the net electric charge, $Q_0$, when $r$ shrink to zero and the lower curve is the net electric charge, $Q$, at $r$ infinity.}
\label{fig.7}
\end{figure}

The electric charge densities, $q=\partial^i E_i$, of these dyon solutions consist of a negative cloud charge distribution concentrated in regions around the origin and along the negative $z$-axis diminishing along the axis as $\frac{1}{r^3}$. The positive charge densities are delta functions distributions at $r=0$ and along the negative $z$-axis. Fig.(\ref{fig.5}) and (\ref{fig.6}) show the electric charge density distribution for the $m=-0.05$ and $m=-0.5$ dyons respectively. The negative electric cloud charge density distribution is in blue and the positive charges are indicated in red. The charge distribution along the negative $z$-axis is greatest when $m=-1/2$ and continuously diminishes to zero as $m$ approaches zero when the dyon's monopole charge changes from half to one. The total electric charge of the dyons can be obtained by Gauss' law,
\begin{eqnarray}
Q &=& \int_{r\rightarrow \infty}E_i\hat{r}_i~r^2\sin\theta~d\theta d\phi,\nonumber\\
&=& 2\pi\sinh\gamma \int\sqrt{m^2 + R(\theta)^2}\sin\theta~d\theta.
\label{eq.24}
\end{eqnarray}
The charge concentrated at the origin is calculated to be
\begin{eqnarray}
Q_0 = 2\pi\sinh\gamma \int\sqrt{(m+1)^2 + R(\theta)^2}\sin\theta~d\theta.
\label{eq.25}
\end{eqnarray}
$Q$ and $Q_0$ can be numerically integrated using Maple 9.5.  The charges of the dyons $Q$ and and $Q_0$ depend on the parameter $m$ and a point plot of $Q$ and $Q_0$ versus $m$ is given in Fig.(\ref{fig.7}). The graph branches out at $m=-1/2$ indicating that there is totally no screening of the positive charge at $r=0$ of the $m=-1/2$ dyon. However as $m$ increases from $-1/2$ to zero, screening of the delta function point source at $r=0$ takes place and when $m=0$, total screening occurs and the net electric charge of the Wu-Yang type dyon is zero.

\section{Comments}
\label{sect.4}
The ansatz (\ref{eq.10}) used to construct the dyon solutions is introduced in the same standard way as in the construction of the Julia-Zee dyons years ago \cite{kn:12}. Hence when the parameter $\gamma$ is zero, the electric field is switched off and only the magnetic field remains and the dyon becomes monopole.

The energy of these dyon solutions are infinite as they possess delta functions electric charge sources at the origin and along the negative $z$-axis. Infinite energy dyons solutions that are complex in the gauge potentials have also been discussed in the literature \cite{kn:14}.

These dyon solutions of one half monopole charge are axially symmetric about the $z$-axis. Their gauge potentials and electromagnetic fields possess a string singularity along the negative $z$-axis. Configurations of one half monopole charge that possess axial symmetry and a semi infinite string singularity both in the gauge potentials and the electromagnetic fields have also been discussed in the literature \cite{kn:11}.

Unlike the monopole solutions of the ansatz (\ref{eq.10}), where the parameter $m$ can take only discrete positive integer values \cite{kn:8}, the dyons of one half monopole charge possess a continuous parameter, $-1/2\leq m < 0$, and as $m$ varies, the properties of the solutions like the electric charges $Q$ and $Q_0$, the electric charge distribution, and the electric and magnetic fields change. However there is a small range of values of $-0.005 \leq m < 0$, for which the monopole charges become indeterminate by Maple 9.5 although we believe that the monopole charge should be one half as long as $m<0$. 

Calculations show that the electric charges that give rise to the $\hat{\theta}_i$ component of the electric field are composed of a negative cloud charge distribution around the negative $z$-axis and a positive delta function charge distribution along the negative $z$-axis which exactly cancelled out the charges of each other at $r$ infinity. 

The electric charges that give rise to the radial component of the electric field are all concentrated at $r=0$ when $m=-1/2$. Hence $Q=Q_0$ for the $m=-1/2$ dyon. However when, $-1/2 < m \leq 0$, there is a negative cloud charge distribution around the origin and this negative cloud charge increases in intensity until it reaches its' maximum value at $m=0$ where it totally screened off the positive delta function source at $r=0$ giving a zero net electric charge for the Wu-Yang type dyon when $m=0$. Hence for the Wu-Yang type dyon, $Q=0$, and $Q_0=4\pi\sinh\gamma$.

\section{Acknowlegements}
The authors would like to thank Universiti Sains Malaysia and the Academy of Sciences Malaysia for the Scientific Advancement Grant Allocation, SAGA, (Account No.: 304/pfizik/653004/A118). The author, Rosy Teh, would also like to thank Prof. Izumi Tsutsui for reading through the manuscript and the Institute of Particle and Nuclear Studies, High Energy Accelerator Research Organization (KEK) Tsukuba, Ibaraki 305-0801, Japan for their hospitality.

\newpage

\section*{FIGURE CAPTIONS}

\end{document}